\documentclass[twocolumn]{aastex631}  

\shorttitle{LAMOST J101356.33+272410.7}
\shortauthors{Yuji He et al.}

\graphicspath{{./}}

\begin{document}

\title{LAMOST J101356.33+272410.7: A Detached White Dwarf-Main-sequence Binary with a Massive White Dwarf Within the Period Gap}

\author[0009-0008-0146-118X]{Yuji He}
\affiliation{Key Laboratory of Optical Astronomy, National Astronomical Observatories, Chinese Academy of Sciences, Beijing 100101, China}
\affiliation{School of Astronomy and Space Science, University of Chinese Academy of Sciences, Beijing 100049, China}

\author[0000-0002-4554-5579]{Hailong Yuan}
\affiliation{Key Laboratory of Optical Astronomy, National Astronomical Observatories, Chinese Academy of Sciences, Beijing 100101, China}

\author[0000-0003-3884-5693]{Zhongrui Bai}
\affiliation{Key Laboratory of Optical Astronomy, National Astronomical Observatories, Chinese Academy of Sciences, Beijing 100101, China}

\author[0009-0000-6595-2537]{Mingkuan Yang}
\affiliation{Key Laboratory of Optical Astronomy, National Astronomical Observatories, Chinese Academy of Sciences, Beijing 100101, China}
\affiliation{School of Astronomy and Space Science, University of Chinese Academy of Sciences, Beijing 100049, China}

\author[0009-0009-6931-2276]{Mengxin Wang}
\affiliation{Key Laboratory of Optical Astronomy, National Astronomical Observatories, Chinese Academy of Sciences, Beijing 100101, China}

\author[0000-0002-6312-4444]{Yiqiao Dong}
\affiliation{Key Laboratory of Optical Astronomy, National Astronomical Observatories, Chinese Academy of Sciences, Beijing 100101, China}

\author[0009-0006-4214-5816]{Xin Huang}
\affiliation{Observatoire de Paris, Paris Sciences et Lettres, CNRS, Place Jules Janssen, F-92195 Meudon, France}

\author[0000-0003-1487-0093]{Ming Zhou}
\affiliation{Key Laboratory of Optical Astronomy, National Astronomical Observatories, Chinese Academy of Sciences, Beijing 100101, China}

\author{Qian Liu}
\affiliation{Key Laboratory of Optical Astronomy, National Astronomical Observatories, Chinese Academy of Sciences, Beijing 100101, China}
\affiliation{School of Astronomy and Space Science, University of Chinese Academy of Sciences, Beijing 100049, China}

\author{Xiaozhen Yang}
\affiliation{Key Laboratory of Optical Astronomy, National Astronomical Observatories, Chinese Academy of Sciences, Beijing 100101, China}
\affiliation{School of Astronomy and Space Science, University of Chinese Academy of Sciences, Beijing 100049, China}

\author{Ganyu Li}
\affiliation{Key Laboratory of Optical Astronomy, National Astronomical Observatories, Chinese Academy of Sciences, Beijing 100101, China}
\affiliation{School of Astronomy and Space Science, University of Chinese Academy of Sciences, Beijing 100049, China}

\author{Ziyue Jiang}
\affiliation{Key Laboratory of Optical Astronomy, National Astronomical Observatories, Chinese Academy of Sciences, Beijing 100101, China}
\affiliation{School of Astronomy and Space Science, University of Chinese Academy of Sciences, Beijing 100049, China}

\author[0000-0002-6617-5300]{Haotong Zhang}
\affiliation{Key Laboratory of Optical Astronomy, National Astronomical Observatories, Chinese Academy of Sciences, Beijing 100101, China}
\affiliation{School of Astronomy and Space Science, University of Chinese Academy of Sciences, Beijing 100049, China}

\begin{abstract}

We report the analysis of the detached eclipsing spectroscopic binary system LAMOST J101356.33+272410.7, which features a massive white dwarf. 
Using LAMOST and SDSS spectra, we determined the stellar parameters and radial velocities of both components. SED fitting of photometric data from GALEX, 2MASS, and Pan-STARRS1 yielded the effective temperatures and photometric radii. Eclipsing analysis of high-speed photometric data from the Liverpool Telescope provided orbital inclination, masses, radii, and related physical parameters.
The white dwarf in this system has a mass of \(1.05 \pm 0.09 \, M_\odot\) and a radius of \(0.0090 \pm 0.0008 \, R_\odot\), while the main-sequence star has a mass of \(0.299 \pm 0.045 \, M_\odot\) and a radius of \(0.286 \pm 0.018 \, R_\odot\). 
Emission lines observed in the spectra indicate the likely presence of stellar magnetic activity in this system.
The relatively cool temperature of the white dwarf suggests that the system could be a post-common-envelope binary (PCEB) that has not undergone mass transfer, while the presence of a massive white dwarf indicates that the system might also represent a detached cataclysmic variable (dCV) crossing the period gap.
We suggest that the system is more likely to be a PCEB, and it is predicted to evolve into a cataclysmic variable and begin mass transfer in approximately 0.27 Gyr.

\end{abstract}




\section{Introduction} \label{sec:intro}  

Cataclysmic variables (CVs) are semi-detached close binary systems in which a white dwarf accretes matter from a low-mass donor star. One prominent feature of these systems is a statistically significant lack of systems with accretion activity in the orbital period range of approximately 2-3 hours, known as the “period gap”. \citet{knigge2006donor} precisely defined the period gap boundaries as \(P_{gap, +}=3.18 \pm 0.04\) hours (upper edge) and \(P_{gap, -}=2.15 \pm 0.03\) hours (lower edge). Recently, \cite{schreiber2024cataclysmic} confirmed the presence of a bimodal feature in the period distribution of the non-polar sample through statistical tests of the SDSS I-IV sample of cataclysmic variables \citep{inight2023catalogue}, indicating a significant orbital period gap. They also determined more precise boundaries for the orbital period gap, which are 147 and 191 minutes, respectively.

To explain the orbital period gap, \citet{rappaport1983} and \citet{spruit1983stellar} proposed the disrupted magnetic braking (DMB) model, which assumes that cataclysmic variables lose angular momentum through two mechanisms: gravitational radiation and stellar wind-driven magnetic braking. For low-mass donors with radiative cores in particular, the efficiency of angular momentum loss due to magnetic braking is much higher than that due to gravitational radiation, causing the orbit to shrink rapidly and leading to a vigorous mass transfer process within the system. This causes the donor to gradually deviate from thermal equilibrium. When the donor's mass reaches \(\thicksim\)0.3\(M_\odot\) (corresponding to an orbital period of about 3 hours), the donor becomes fully convective.
At this stage, spin-up of the donor leads to a more complex magnetic field dominated by high-order multipole components rather than a simple dipolar configuration \citep{garraffo2018magnetic}. This complexity reduces the number of open magnetic field lines, significantly decreasing angular momentum loss through magnetic braking. Consequently, the mass transfer rate drastically decreases, the donor’s radius shrinks, and it retreats within its Roche lobe, forming a detached WD + MS binary, also referred to as a detached CV (dCV). 
 CVs pass through the period gap in the form of dCVs, during which they lose angular momentum via gravitational radiation and gradually move closer until \(P_{\text{orb}} \thicksim 2\,\text{h}\).

Over the past few decades, the DMB theory has become the standard model for the evolution of CVs. However, several inconsistencies between the model and observations persist. 
First, the theoretically predicted minimum orbital period of 65-70 minutes \citep[][]{kolb1993,kolb1999brown,howell2001exploration,goliasch2015population, kalomeni2016evolution} is shorter than the observed range of 76-82 minutes \citep{knigge2006donor,knigge2011evolution,mcallister2019evolutionary}. 
Second, the standard model predicts that \( 90\%-99\% \) of CV systems should be located below the period gap \citep[e.g.][]{kolb1993,howell2001exploration,goliasch2015population}, whereas observations indicate that \(83 \pm 6\%\) of systems are below the gap and \(17 \pm 6\%\) are above it \citep{pala2020volume}. 
Third, the theoretical prediction of CV spatial density is 1-2 orders of magnitude higher than observed values \citep{ritter1986mass,pretorius2012space}. 
While the first two issues have been alleviated by the Sloan Digital Sky Survey (SDSS), which has facilitated the detection of fainter systems. This advancement has, in turn, revealed a new challenge: the white dwarf mass discrepancy. Specifically, the standard model predicts an average white dwarf mass of \(\sim 0.6 M_\odot\) \citep{kepler2017white}, significantly lower than the observed mean mass of \(\sim 0.82 M_\odot\) \citep{pala2020volume} in CVs. 

To address the potential white dwarf mass problem, \citet{king1995consequential} and \citet{schreiber2015three} introduced the concept of consequential angular momentum loss (CAML)-a theoretical mechanism describing angular momentum loss caused by mass ejection during nova eruptions. They proposed two possible models: the classical non-conservative CAML model (cCAML) and the empirical CAML model (eCAML).
The eCAML model not only contributes to resolving the white dwarf mass issue but also unexpectedly improves the problem of space density of CVs and explains the existence of single low-mass WDs \citep{zorotovic2020cataclysmic}. This is because the eCAML model predicts a greater proportion of CV mergers compared to previous models, particularly in systems with low-mass WDs. Furthermore, the test results of the eCAML model align more closely with the observational characteristics of CVs.


Observationally, \citet{davis2008many} suggested that observing detached WD + MS binaries within the orbital period gap could provide direct validation for both the existence of the gap and the disrupted magnetic braking theory.
In particular, the donor star's radius \citep{knigge2011evolution} and the white dwarf's effective temperature \citep{townsley2009cataclysmic} serve as sensitive tracers of the mass-transfer rate in cataclysmic variables. The donor's radius is primarily determined by its mass-loss rate, as this influences its thermal equilibrium, while the accretion of material heats the white dwarf. Both parameters provide valuable insights into the system's mass-transfer rate, thereby enhancing our understanding of the CV's evolutionary history.
Observations of eclipsing systems allow precise determination of the orbital inclination, enabling accurate measurements of stellar mass, radius, and other physical parameters. If these systems are also spectroscopic binaries, highly precise orbital parameters can be obtained, offering comprehensive observational data. Such systems are of high analytical value and can further validate the aforementioned theories.

J101356.33+272410.7 (hereafter J1013+2724) exemplifies such a binary system, with an orbital period of 0.129040379 {\rm days}, as reported by \citet{parsons201514}. They also found that the white dwarf in this system has an effective temperature of approximately \(15601 \, {\rm K}\), while the companion main-sequence star is classified as M4-type. Using SDSS spectral data, \citet{rebassa2012post} derived an effective temperature of \(T_{\rm eff} = 16526 \pm 277 \, {\rm K}\) and a surface gravity of \(\log g = 8.8 \pm 0.044 \, {\rm dex}\) for the white dwarf in J1013+2724. The companion main-sequence star was also identified as M4-type. By applying the tables of \citet{bergeron1995photometric}, the white dwarf's mass and radius were determined to be \(1.1 \pm 0.023 \, M_\odot\) and \(0.00698 \pm 0.00027 \, R_\odot\), respectively. The main-sequence star's mass and radius were estimated as \(0.319 \pm 0.09 \, M_\odot\) and \(0.326 \pm 0.096 \, R_\odot\) using the empirical Sp-M-R relation from \citet{rebassa2007post}.

We conducted a more detailed study of this system. In Section \ref{sec:observations}, the observational data of J1013+2724 are introduced. Section \ref{sec:data analysis} provides a detailed description of the analysis process for the spectroscopic and photometric data. In Section \ref{sec:result}, we summarize and discuss the results of this analysis, including the system's evolution history and potential future evolution.


\section{Observations} \label{sec:observations}  

J1013+2724 is an eclipsing spectroscopic binary consisting of a massive white dwarf and an M-type main-sequence star. For clarity, we refer to the main-sequence star as Star 1 and the white dwarf as Star 2. All timestamps were converted to the Barycentric Julian Date (BJD) in the Barycentric Dynamical Time (TDB) standard \citep{eastman2010achieving} to eliminate time discrepancies arising from Earth's motion, thereby enhancing the precision of astronomical observations. This section describes the spectroscopic and photometric data utilized in this study.


\subsection{Spectroscopic Observation} \label{subsec:spec}  

The Large Sky Area Multi-Object Fiber Spectroscopic Telescope (LAMOST, \citealt{cui2012large}) is an active reflecting Schmidt telescope capable of simultaneously acquiring 4,000 spectra. It has an effective aperture of 4 meters and a 5-degree field of view. In its low-resolution mode, LAMOST provides a spectral resolution of approximately 1800, spanning wavelengths from 3700 to 9000 \AA.
Between September 2015 and March 2024, J1013+2724 was observed 10 times using LAMOST in low-resolution mode. Each observation had an exposure time of 1800 seconds, except for two conducted in 2024, which used exposure times of 1500 and 1100 seconds, respectively. The raw CCD images were processed with the LAMOST two-dimensional pipeline \citep{bai2021first}, which includes bias and dark subtraction, flat-field correction, spectral extraction, sky background subtraction, cosmic ray removal \citep{bai2016cosmic}, and wavelength calibration \citep{bai2017sky}.

Listed in the spectroscopic catalog of white dwarf-main-sequence (WDMS) binaries \citep{rebassa2013white} from the Sloan Digital Sky Survey \citep[SDSS,][]{york2000sloan}, J1013+2724 was observed four times on February 21, 2009. These spectra span a wavelength range of 3800-9200 \AA, with a resolution of 1800.


\begin{table*}[t]
\centering
\caption{Details of observational data from the Liverpool Telescope}
    \begin{tabular}{cccccc}
    \hline
    Start of Observation & End of Observation & EXPTIME & Instrument & Filter & Number of exposures\\ \hline
    2013-04-09 00:15:35.152 & 2013-04-09 00:52:13.009 & 10s & RISE & og515+kg5 & 220 \\ 
    2014-03-08 23:11:29.748 & 2014-03-08 23:58:19.784& 10s & RISE & og515+kg5 & 280\\ 
    2015-01-24 00:24:44.061 & 2015-01-24 01:11:34.121& 10s & RISE & og515+kg5 & 280\\ \hline
    \end{tabular}
\label{tab:lt_data}
\end{table*}

\subsection{Light Curves} \label{sec:lc}  

High-speed photometric observations of J1013+2724 were obtained from the 2-meter Liverpool Telescope \citep[LT,][]{steele2004liverpool} through its data archive website\footnote{\url{http://telescope.livjm.ac.uk/}}. Observations were conducted from 2012 to 2015 using the high-speed RISE camera and a single wide-band V+R filter. The raw images downloaded had already undergone basic instrumental corrections, such as bias subtraction, dark current removal, and flat-field correction. We then used AstroImageJ\footnote{\url{https://www.astro.louisville.edu/software/astroimagej/}} to perform differential photometry and obtained the corresponding light curves. The resulting light curve is shown in Figure \ref{fig:lt}.
Photometric data from 2012, which included incomplete eclipses and insufficient information to characterize the binary system, were excluded from the analysis. Instead, we utilized high-quality data collected between 2013 and 2015 with 10-second exposure times. Table \ref{tab:lt_data} summarizes these observations, including exposure times and observational settings.

The Zwicky Transient Facility \citep[ZTF,][]{masci2018zwicky}, a ground-based optical time-domain survey using the 48-inch Schmidt telescope at Palomar Observatory, provided publicly available photometric data for J1013+2724. The ZTF light curve data were downloaded from the Infrared Science Archive (IRSA)\footnote{\url{https://irsa.ipac.caltech.edu/Missions/ztf.html}}. We focused on the r- and g-band observations collected between 2018 and 2023, each with a consistent exposure time of 30 seconds. 
Using the eleanor tool\footnote{\url{https://adina.feinste.in/eleanor/}} \citep{feinstein2019eleanor}, we downloaded full-frame images (FFIs) of J1013+2724 observed by the Transiting Exoplanet Survey Satellite (TESS) in 2022, corresponding to Sector 48 with an exposure time of 10 minutes. Then, the FFIs were processed to extract light curves, and the flux was converted to absolute magnitudes based on the TESS magnitude of \(15.600 \pm 0.026 \, \text{mag}\), retrieved from the Mikulski Archive for Space Telescopes (MAST) using the \texttt{astroquery.mast} module\footnote{\url{https://astroquery.readthedocs.io/en/latest/mast/mast.html}}.
As shown in Figure \ref{fig:rv_fit}, the ZTF and TESS light curves complement the high-cadence Liverpool Telescope data by providing supplementary information for phases outside of the eclipses, as well as additional photometric information in two filters.

Photometric data from the Catalina Sky Survey \citep[CSS][]{drake2009first}, covering 2005 to 2012, were included to extend the observational baseline. By performing Lomb-Scargle periodogram analysis on the Catalina, ZTF, and TESS datasets, we determined the orbital period of the system to be 0.1290403131 \text{days}. Notably, the orbital period showed no significant variation over the past two decades, demonstrating its long-term stability.
The orbital phase is defined using the following ephemeris:
\begin{equation}
	T(\phi=0)= 2458425.8493(5)\text{BJD}+0.1290403(1)\times E
\end{equation}
where  \(\phi=0\) corresponds to the time when the white dwarf is in front of the main-sequence star, and  \(\phi=0.5\) marks the midpoint of the eclipse, when the main-sequence star completely obscures the white dwarf. E is the orbital cycle number.

\subsection{Multi-band Photometry}\label{subsec:pho}  

The SDSS photometry, based on single exposures, shows phase-dependent variability and is further compromised by the source’s position near the edge of the image, leading to inaccuracies; therefore, it was not used in our analysis.
In the ultraviolet band, the binary system was observed by GALEX\footnote{\url{http://www.galex.caltech.edu/index.html}}, providing data in the FUV and NUV bands. In the near-infrared and optical bands,  the Two Micron All Sky Survey\footnote{\url{https://irsa.ipac.caltech.edu/Missions/2mass.html}} (2MASS) and the Panoramic Survey Telescope and Rapid Response System 1\footnote{\url{https://www.mpia.de/en/research/collaborations/pan-starrs1}} (Pan-STARRS1) conducted observations of this source. These photometric data, listed in Table \ref{tab:sed_data}, were subsequently used in our analysis. They effectively characterize both the hot and cool components of the binary system.

\begin{table}[t]
\centering
\caption{Multiband Photometry Measurements}\label{tab:sed_data}
\begin{tabular}{p{1.5cm} p{1cm} p{2.5cm}} 
\hline
\multicolumn{1}{l}{Telescope} & \multicolumn{1}{l}{Band} & \multicolumn{1}{l}{Magnitude} \\
\hline
    GALEX &FUV & 18.033 $\pm$ 0.071 \\ 
    GALEX &NUV & 17.956 $\pm$ 0.047 \\ 
    2MASS &J & 13.825 $\pm$ 0.025 \\ 
    2MASS &H & 13.255 $\pm$ 0.029 \\ 
    2MASS &K & 12.947 $\pm$ 0.028 \\ 
    PS1 &g & 17.5916 $\pm$ 0.0078 \\ 
    PS1 &r & 17.2319 $\pm$ 0.0153 \\ 
    PS1 &i & 16.0572 $\pm$ 0.0069 \\ 
    PS1 &z & 15.5512 $\pm$ 0.0330 \\ 
    PS1 &y & 15.1920 $\pm$ 0.0166 \\
\hline
\end{tabular}
\end{table}


\section{Data analysis} \label{sec:data analysis}  


This section details the data analysis process for J1013+2724. In Section \ref{subsec:spec_para}, the stellar parameters and radial velocities of both components are derived by fitting their spectra. Section \ref{subsec:rv_fit} addresses the fitting of radial velocity data to determine orbital parameters, including the period, radial velocity semi-amplitudes, systemic velocity, mass ratio, and gravitational redshift. In Section \ref{subsec:sed}, the photometric radii of both components are estimated through SED fitting. Finally, Section \ref{subsec:eclip} presents the refinement of orbital inclination, stellar masses, and radii by modeling high-cadence photometric data with a simplified orbital model.

\begin{figure}[t]
\centering
\includegraphics[width=\columnwidth]{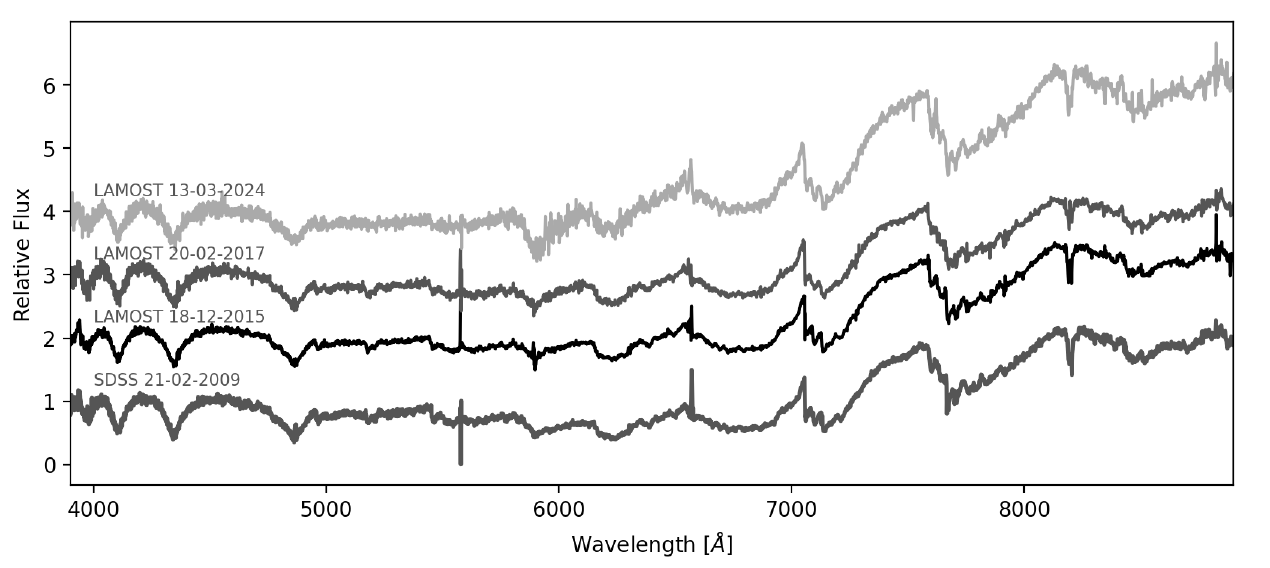}
\caption{The combined spectra of J1013+2724 from LAMOST and SDSS.\label{fig:cospec}}
\end{figure}

\subsection{Spectral Parameters}\label{subsec:spec_para}  

The combined spectra of J1013+2724 obtained from LAMOST and SDSS are shown in Figure \ref{fig:cospec}. In the blue end of the spectrum, the flux is dominated by the white dwarf, while the red end is primarily contributed by the main-sequence star.
The spectra clearly reveal the characteristics of both stars, with a series of distinct hydrogen Balmer lines also prominently visible.

Based on the characteristics described above, we first decomposed the two components from the observed spectra. Then, we determined the stellar parameters of the white dwarf and main-sequence star components using the appropriate templates. The Koester DA template \citep{koester2010white}, was used to determine the white dwarf parameters after interpolation. This grid covers the range of \(T_{\text{eff}} = 6000-100,000 K\) and \(\log{g} = 5.0-9.5\). For the main-sequence star, we used the BT-Settl template \citep{allard2013bt}. Prior to applying these templates, we performed a convolution operation to degrade them to the same resolution as the observed spectra (R = 1800).

\begin{figure*}[t]
\includegraphics[width=\textwidth]{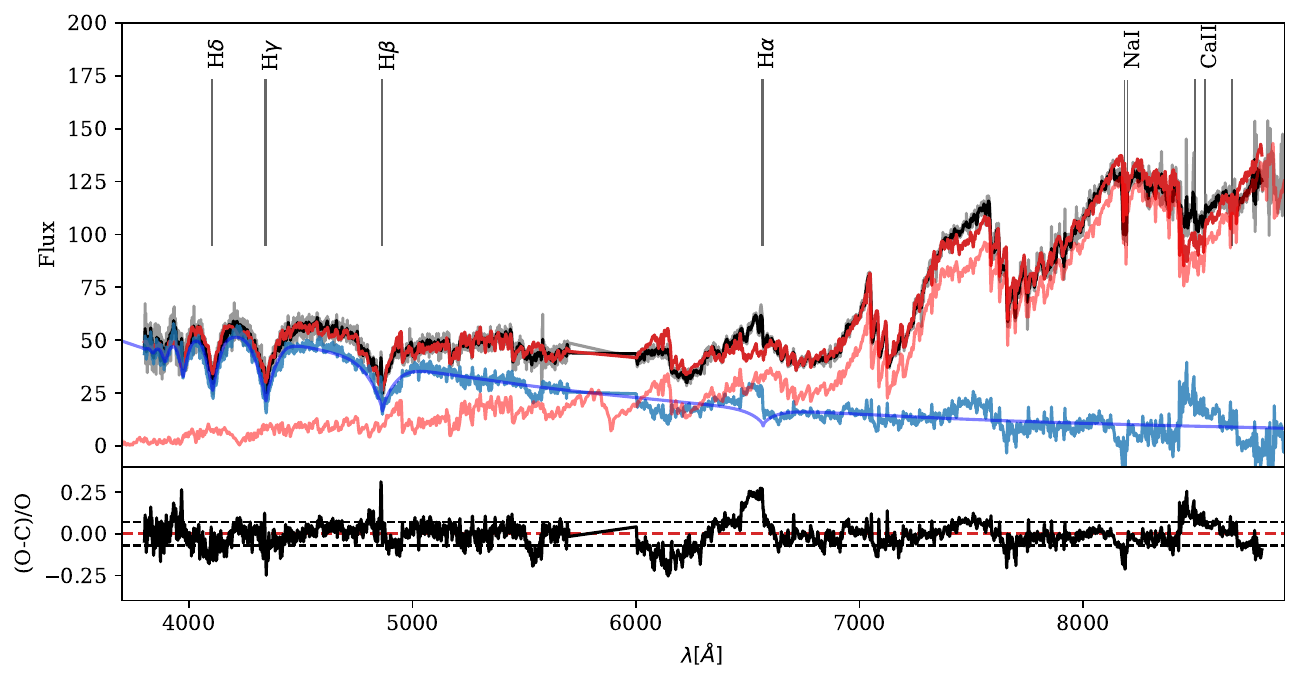}
\caption{\textit{Top}: the results of spectral decomposition and the subtraction of the main-sequence star template.
The black curve represents the smoothed observed spectra (gray), while the red curve is the best-fit synthetic spectrum, composed of the KoesterDA template for the white dwarf (dark blue) and the main-sequence star template (light red). The light blue curve represents the clean white dwarf observed spectra, with the main-sequence star's contribution removed. The data in the 5700-6000 Å range was trimmed to improve the quality of the observed spectra.
\textit{Bottom}: the residuals between the observed spectra and the template-synthesized spectra.}
\label{fig:decom_spec}
\end{figure*}

In the first step, we applied a double Gaussian function to fit the Na I 8183/8194 absorption lines, yielding a reasonable estimate of the main-sequence star's radial velocity from the Doppler shift. To accurately fit the white dwarf spectrum and obtain reliable parameters and radial velocities, it is crucial to subtract the contribution of the main-sequence star from all individual spectra \citep{parsons2017testing}. A straightforward approach involves subtracting the spectrum observed during the primary eclipse, where the white dwarf is entirely obscured by the main-sequence star, and the flux originates solely from the latter. Unfortunately, no in-eclipse spectra are available for this system. Instead, we subtract a best-fit main-sequence star template (with parameters $T_{\text{eff}} = 3307$ K, $\log{g} = 4.65$, and [Fe/H] = 0.29), shifted according to the radial velocities obtained in the first step.
This main-sequence star template was derived by minimizing the \(\chi^2\) value between the red-arm spectrum (6750-8700 \AA) and the interpolated BT-Settl templates. Figure \ref{fig:decom_spec} illustrates an example of the results obtained through these procedures.

Following the above procedure, the absorption features of the main-sequence star were largely removed. However, some emission lines, such as those at the cores of the \(H\beta\) and \(H\gamma\) absorption lines, remain and vary with orbital phase. To minimize their impact, we restricted the wavelength range for subsequent fitting to the blue end, dominated by the white dwarf's spectrum, specifically below 5700 \AA.

The \textit{corv} package \citep{arseneau2024measuring} in Python was utilized to derive the radial velocity and stellar parameters of the white dwarf from its subtracted spectrum. Instead of using Voigt profile fitting, we analyzed regions near the Balmer absorption lines with DA model templates from \citet{koester2010white}, as Stark broadening in high-$\log{g}$ stars often degrades measurement accuracy. The H$\beta$ line was excluded from fitting due to its large equivalent width and phase-dependent emission at the line center, which could introduce significant errors. Similarly, regions with low signal-to-noise ratios below 4000 Å were not used. Only the H$\gamma$ and H$\delta$ lines were included in the radial velocity calculations, with results summarized in Table \ref{tab:rv}. The stellar parameters of the white dwarf, derived from our fitting, are $T_{\rm eff} = 15661 \pm 973 \, {\rm K}$ and $\log g = 8.71 \pm 0.10 \, {\rm dex}$.

\begin{table*}[!t]
\centering
\caption{Spectroscopic Observations and Estimated Parameters of J1013+2724}\label{tab:rv}
\begin{tabular}{llllllll}
\hline
\hline
Date & BJD & $RV_{MS}$ & $RV_{WD}$ & S/N & Exptime & Telescope \\
day-month-year & day & $km\,s^{-1}$ & $km\,s^{-1}$ &  & s & \\
\hline
    18-12-2015 & 2457375.318 & -214.67 $\pm$ 4.54 & 172.02 $\pm$ 29.90 & 22 & 1800 & LAMOST \\ 
    18-12-2015 & 2457375.341 & 210.41 $\pm$ 4.95 & 42.39 $\pm$ 39.66 & 20 & 1800 & LAMOST \\ 
    18-12-2015 & 2457375.365 & 356.70 $\pm$ 4.30 & -0.11 $\pm$ 29.53 & 20 & 1800 & LAMOST \\ 
    18-12-2015 & 2457375.388 & 221.79 $\pm$ 7.73 & -0.45 $\pm$ 28.61 & 19 & 1800 & LAMOST \\ 
    18-12-2015 & 2457375.411 & -216.19 $\pm$ 4.66 & 168.97 $\pm$ 26.70 & 20 & 1800 & LAMOST \\ 
    20-02-2017 & 2457805.138 & -318.80 $\pm$ 11.30 & 190.93 $\pm$ 30.71 & 15 & 1800 & LAMOST \\ 
    20-02-2017 & 2457805.161 & -82.53 $\pm$ 16.50 & 46.03 $\pm$ 59.66 & 10 & 1800 & LAMOST \\ 
    20-02-2017 & 2457805.184 & 306.24 $\pm$ 8.45 & 44.47 $\pm$ 28.28 & 15 & 1800 & LAMOST \\ 
    13-03-2024 & 2460383.185 & 199.17 $\pm$ 11.18 & -56.91 $\pm$ 33.87 & 20 & 1500 & LAMOST \\ 
    13-03-2024 & 2460383.204 & -145.92 $\pm$ 9.71 & 167.96 $\pm$ 34.65 & 20 & 1100 & LAMOST \\ 
    21-02-2009 & 2454883.690 & -142.06 $\pm$ 25.06 & 186.15 $\pm$ 48.19 & 11 & 1800 & SDSS \\ 
    21-02-2009 & 2454883.713 & 323.32 $\pm$ 9.25 & -12.38 $\pm$ 27.44 & 9 & 1800 & SDSS \\ 
    21-02-2009 & 2454883.740 & 276.85 $\pm$ 11.19 & 16.43 $\pm$ 24.78 & 13 & 1800 & SDSS \\ 
    21-02-2009 & 2454883.763 & -101.82 $\pm$ 22.33 & 238.21 $\pm$ 24.27 & 20 & 1800 & SDSS \\ 
\hline
\end{tabular}
\end{table*}

\subsection{Orbital Solution}\label{subsec:rv_fit}  

The orbital period of J1013+2724 is 185.82 minutes, while the exposure time for most observed spectra is 30 minutes, corresponding to a phase coverage of 0.167. To account for the impact of phase smearing on the radial velocity amplitude \citep{yuan2023elm}, we applied the following correction for circular orbits:
\begin{equation}
	V(\phi)-V_0=(V_{obs}-V_0)\cdot2\pi\delta\phi/sin(2\pi\delta\phi)  \label{eq:smearing_effect}  
\end{equation}
where $V(\phi)$ is the true velocity at phase $\phi$, \(V_0\) is the systemic velocity, \(V_{\text{obs}}\) is the observed velocity, and $\delta\phi$ represents half of the phase covered during the exposure. Radial velocities for both components were iteratively corrected for this effect.


Orbital parameters were derived by fitting the radial velocity data using the equation:
\begin{equation}\label{eq:rv}
	V(t)=\gamma+K \, sin\left[ \frac{2\pi(t-t_0)}{P_{orb}} \right]   
\end{equation}
where \(K\) is the radial velocity semi-amplitude, \(t_0\) is the mid-eclipse time, \(P_{\text{orb}}\) is the orbital period, and \(\gamma\) is the systemic velocity. The mass ratio, \(q = M_2/M_1 = K_1/K_2\), was derived from the ratio of the semi-amplitudes, while differences in systemic velocities arise from the gravitational redshift effect.



Gravitational redshift occurs as photons lose energy while escaping a gravitational field. For a main-sequence star, this effect is negligible due to its weaker gravitational field. However, for a white dwarf with a stronger gravitational field, photons emitted from its surface must overcome a deep gravitational well, leading to measurable spectral line shifts. Consequently, the systemic velocities \(\gamma_1\) and \(\gamma_2\) is different.

\begin{figure}[t!]
	\centering
	\includegraphics[width=\columnwidth]{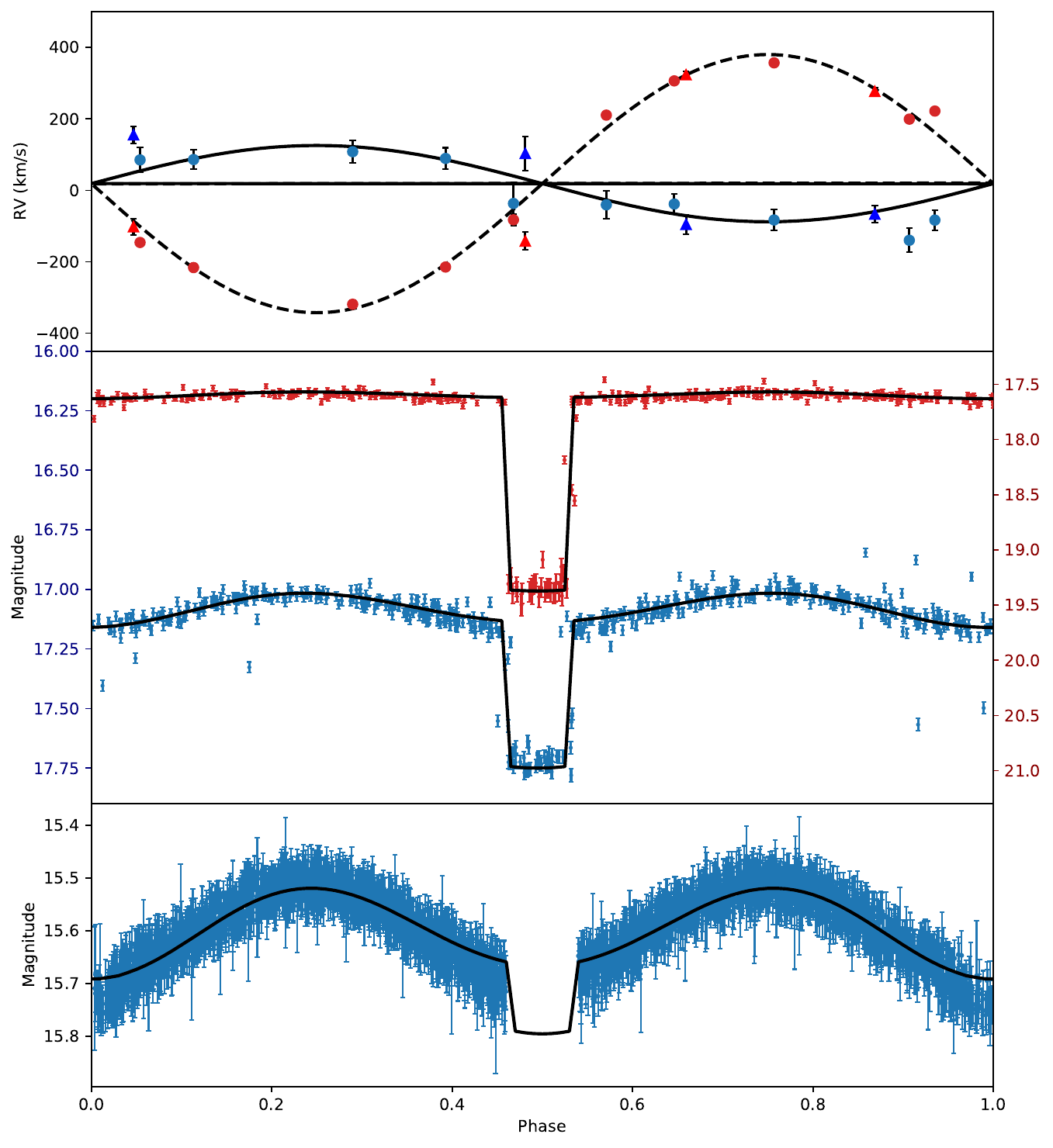} 
	\caption{\textit{Top}: The radial velocity curve of J1013+2724, where the red circles (LAMOST) and triangles (SDSS) represent the radial velocity data of the main-sequence star, while the blue circles (LAMOST) and triangles (SDSS) represent the radial velocity data of the white dwarf. The solid black line and dashed black line show the best-fit curves obtained using Equation \ref{eq:rv} for the two components. (Note: The velocity of the white dwarf has been corrected for gravitational redshift.)
	\textit{Bottom}: The light curves of J1013+2724 in the ZTF r-band (red points) and g-band (blue points), with the solid black line and dashed black line representing the theoretical light curves generated by the Wilson-Devinney code.}
	\label{fig:rv_fit}
\end{figure}

Figure \ref{fig:rv_fit} shows the radial velocity fitting curve, and Table \ref{tab:results} lists the best-fit parameters. From the radial velocity semi-amplitudes, the mass ratio of the white dwarf to the main-sequence star is derived as \(q = 3.4 \, \pm  \, 0.3\), and the gravitational redshift is \(v_g = 82.9 \pm 10.2 \, \text{km/s}\).

\subsection{Spectral Energy Distribution Fitting} \label{subsec:sed}  


The spectral energy distribution (SED) describes the flux at different wavelengths, shaped by the intrinsic properties of the star (e.g., temperature and metallicity) and external factors.A significant portion of the ultraviolet and visible light emitted by the star is absorbed by the interstellar medium, which consists of dust and gas, and is then re-emitted at longer wavelengths. Moreover, extinction becomes more severe with increasing distance, making the star appear fainter.

Based on GAIA DR3 data \citep{gaia2022vizier}, J1013+2724 has a parallax of \(\varpi = 7.4689 \pm 0.0644\) mas, corresponding to a distance of \(D = 133.89 \pm 1.15\) pc. The distance modulus (DM) is calculated as \(5.63 \pm 0.02\). Using the 3D dust map by \citet{green2018galactic}, the extinction coefficient \(\alpha\) is 0.0101, resulting in \(E(B-V) = 0.009 \pm 0.003\).

Because SED fitting is highly sensitive to the effective temperature but less sensitive to surface gravity \(\log g\) and metallicity, it is reasonable to fix the latter two parameters using the results from Section \ref{subsec:spec_para}. This method improves the stability and reliability of the SED fitting results.
Therefore, the relationship between the apparent flux  \( f_\nu \)  and the surface flux \( F_\nu \)  of the binary system at a given wavelength and effective temperature is expressed as
\begin{equation}
f_\nu=\left(\frac{R_1}{D} \right)^2  
\left[ F_{\nu, 1}(T_{eff,1})+ 
a\cdot F_{\nu, 2}(T_{eff,2}) \right]
\end{equation}
 In this equation, \( a = \left( R_2 / R_1 \right)^2 \) represents the squared ratio of the radii of the two stars, where  \( R_1 \) and \( R_2 \) denote the radii of the main-sequence star and the white dwarf, respectively. \( D \) is the distance to the system.

\begin{figure}[t]
    \centering
    \begin{minipage}[b]{\columnwidth}
        \centering
        \includegraphics[width=\columnwidth]{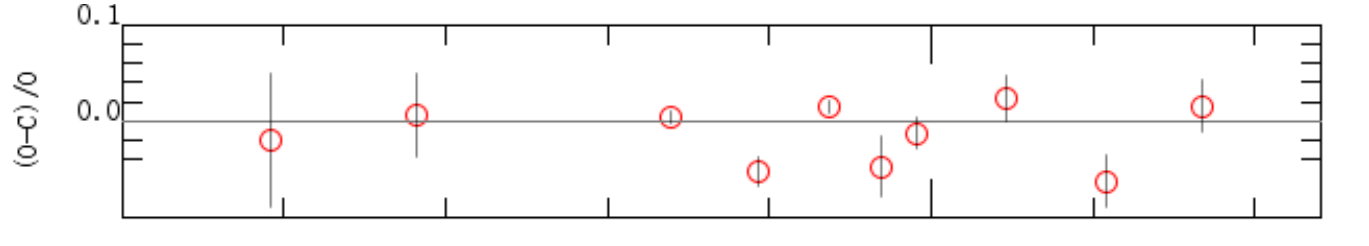}
        \vspace{-\baselineskip} 
        \includegraphics[width=\columnwidth]{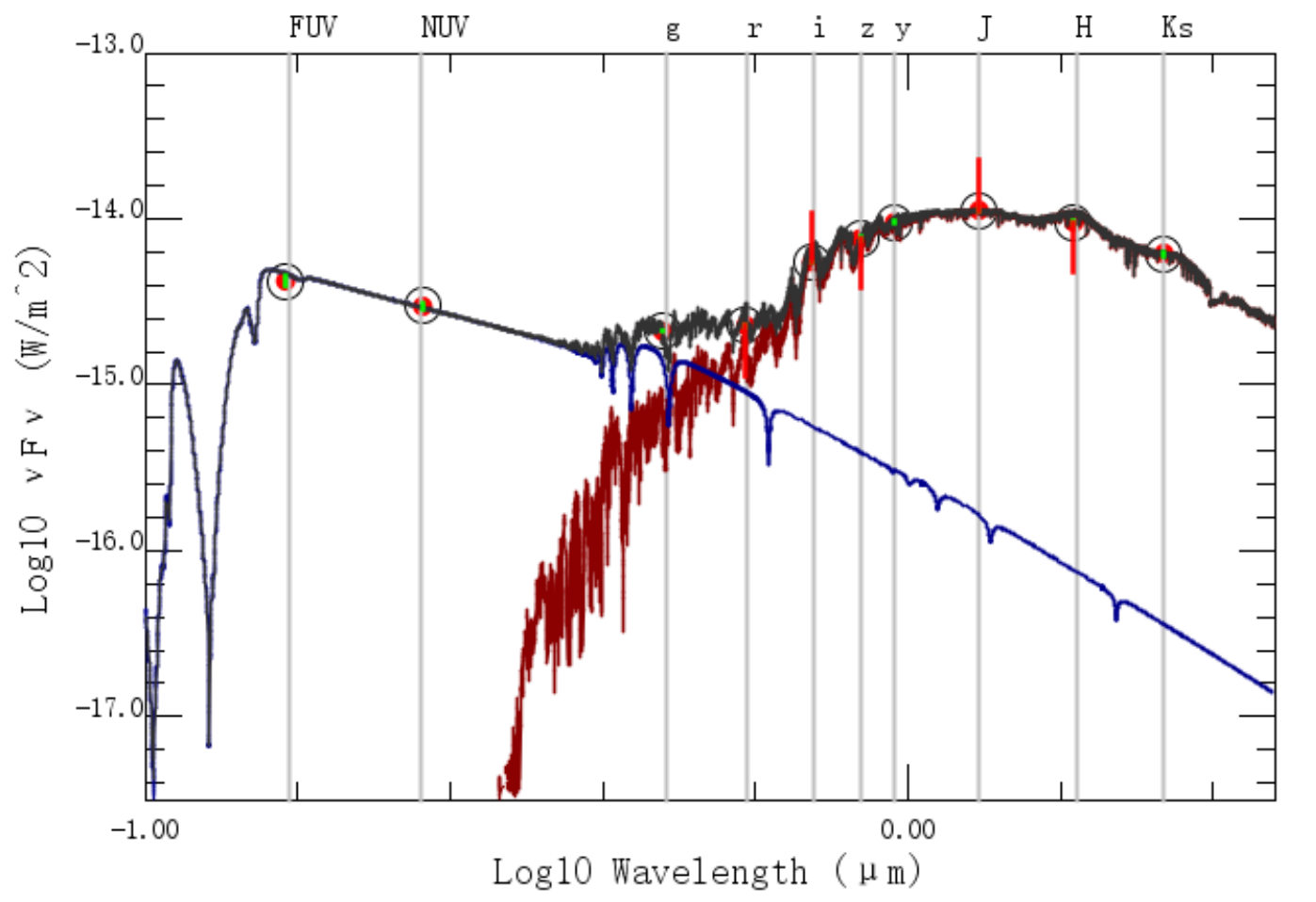}
    \end{minipage}
    \caption{The SED fitting of J1013+2724 with two components. The red dots denote observations used for fitting, including GALEX, 2MASS, and PS1 observations. The blue curve represents the Koester DA spectral template with parameters of \(T_{eff}=15606\) K and \(log{g}=8.71\); the red curve represents the BT-Settl spectral template with parameters of \(T_{eff}=3196\) K, \(log{g}=4.65\), and [Fe/H]=0.3. The gray curve shows the composite spectral template. The residuals are calculated as \((Observed - Calculated)/Observed\).}
    \label{fig:sed_fitting}
\end{figure}

The photometric data were simultaneously fitted using the Koester DA \citep{koester2010white} and BT-Settl \citep{allard2013bt} stellar atmosphere models. To achieve this, model parameters were systematically adjusted to reproduce the observed SED, with a particular focus on capturing the luminosity ratio of the two components. During the fitting process, the free parameters considered were \(R_1\), a, and the effective temperatures $T_{\rm eff}$ of both stars. To assess the consistency between the model and the observed data, a chi-squared minimization method was employed. This approach allowed for the determination of the best-fitting effective temperature and radius for J1013+2724. The uncertainties of the parameters were derived through Markov Chain Monte Carlo (MCMC) simulations to provide robust error estimates. Finally, we determined the best-fitting parameters for J1013+2724, with the resulting temperatures and photometric radii presented in Table \ref{tab:results} and the corresponding fitting plot shown in Figure \ref{fig:sed_fitting}.

\begin{table}
\centering
\setlength{\belowcaptionskip}{0pt} 
\caption{Summaries of Parameters for J1013+2724}\label{tab:results}
\begin{tabular}{llll}
\hline
\textbf{Name} & \textbf{Value} & \textbf{Unit} & \textbf{Description} \\
\hline
R.A.         & 153.48472285    & degrees    & \\
Decl.        & +27.40306405     & degrees    & \\
\hline
$T_{\rm eff,1}$ & 3307.5 $\pm$ 39.5   & K  & Spectral Fitting\\
log $g$      & 4.65 $\pm$ 0.09   & dex  & \\
{[Fe/H]}     & 0.29 $\pm$ 0.04   & dex  & \\
$T_{\rm eff,2}$ &  15661 $\pm$ 973   & K     &   \\
log $g$      & 8.71 $\pm$ 0.10  & dex  & \\ 
\hline
P            & 0.1290403(1)      & days  & Orbital Solution \\ 
$T_0$       &  2458425.8493(5) & days       &  \\
$K_1$       & 361.2 $\pm$ 11.2 & km s$^{-1}$ &  \\ 
$K_2$       & 106.7 $\pm$ 12.2 & km s$^{-1}$ &  \\ 
$\gamma_1$  & 18.8 $\pm$ 7.9 & km s$^{-1}$ & \\ 
$\gamma_2$  & 101.7 $\pm$ 6.6  & km s$^{-1}$ & \\
\hline
Plx          & 7.4689 $\pm$ 0.0644 & mas      & SED Fitting \\
D            & 133.9 $\pm$ 1.2 & pc       & \\
E(B-V)      & 0.0194 $\pm$ 0.006 & mag       & \\
$T_{\rm eff, 1}$ & 3196 $^{+83}_{-80}$   & K          & \\
$T_{\rm eff, 2}$ & 15606 $^{+394}_{-353}$   & K          & \\
$R_1$            & 0.275$^{+0.022}_{-0.019}$ & R$_{\odot}$ & \\
$R_2$           & 0.0081$^{+0.0003}_{-0.0001}$ & R$_{\odot}$ & \\
\hline
Inc           & 83.2 $^{+2.4}_{-1.9}$  & degrees  &  Eclipse Modeling\\
SMA          &  1.19 $^{+0.04}_{-0.04}$  & R$_{\odot}$  &  \\
q           & 3.5 $^{+0.3}_{-0.3}$  &   & \\
$K_1$       & 359.86 $^{+9.61}_{-9.43}$  & km s$^{-1}$ &  \\
$M_1$        & 0.299$^{+0.045}_{-0.037}$ & M$_{\odot}$ & \\ 
$R_1$        & 0.286$^{+0.018}_{-0.017}$ & R$_{\odot}$ & \\
$M_2$        & 1.05$^{+0.09}_{-0.09}$ & M$_{\odot}$ & \\  
$R_2$        & 0.00897$^{+0.00077}_{-0.00075}$ & R$_{\odot}$ & \\
\hline
$A_g$   &  0.035 $\pm$ 0.012   & mag   & Extinction\\
$A_r$   &  0.025 $\pm$ 0.008   & mag   & \\
$A_{\text{TESS}}$   &  0.020 $\pm$ 0.007   & mag   & \\
$R_1/R_{L, 1}$ & 88.5 $\pm$ 6.9  & \% & Filling factor\\
\hline
\end{tabular}
\end{table}


\subsection{Analyzing eclipses}\label{subsec:eclip}  

This section details the application of the direct method to analyze the LT light curve of the eclipsing system, along with the results obtained. The parameter distribution of the binary system is determined by combining this analysis with Markov Chain Monte Carlo simulations. 

\subsubsection{Fitting the LT light curve}\label{subsubsec:lt_fit}  

Eclipsing system light curves offer additional constraints, enabling precise determination of stellar parameters such as radius, mass ratio, and orbital inclination without relying on atmospheric models. For example, the durations of ingress and egress, as well as the eclipse width, constrain the radii of the both components, while the eclipse depth depends on the white dwarf's radius and relative temperature. Therefore, the LT light curve was first fitted to extract characteristic features, facilitating the subsequent determination of orbital and binary parameters consistent with these features.

Following the method of \citet{bours2014testing}, we modified the approach by fitting the ingress and egress of the white dwarf's eclipse with two symmetric sigmoid functions centered on the eclipse midpoint. Furthermore, the method was refined to better suit our data by fitting the out-of-eclipse flux with a horizontal line:
\begin{equation}
f=\frac{d}{1+e^{k\cdot(\theta+w_1-c)}}+\frac{d}{1+e^{-k\cdot(\theta-w_1-c)}}+s
\label{eq:eclip}
\end{equation}
Here, $\theta$ and $f$ denote the phase and flux of the light curve, respectively, while $d$, $k$, $w_1$, $c$, and $s$ are the fitting coefficients. The coefficients have specific interpretations based on the sigmoid function's properties: $d$ represents the eclipse depth, $c$ is the phase at mid-eclipse, $w_1$ quantifies the phase difference between ingress/egress midpoints and the eclipse midpoint, and $s$ accounts for the overall flux offset. The steepness of ingress and egress, represented by $k$, reflects the rate of flux change during the white dwarf's eclipse.

\begin{figure}[t]
	\includegraphics[width=\columnwidth]{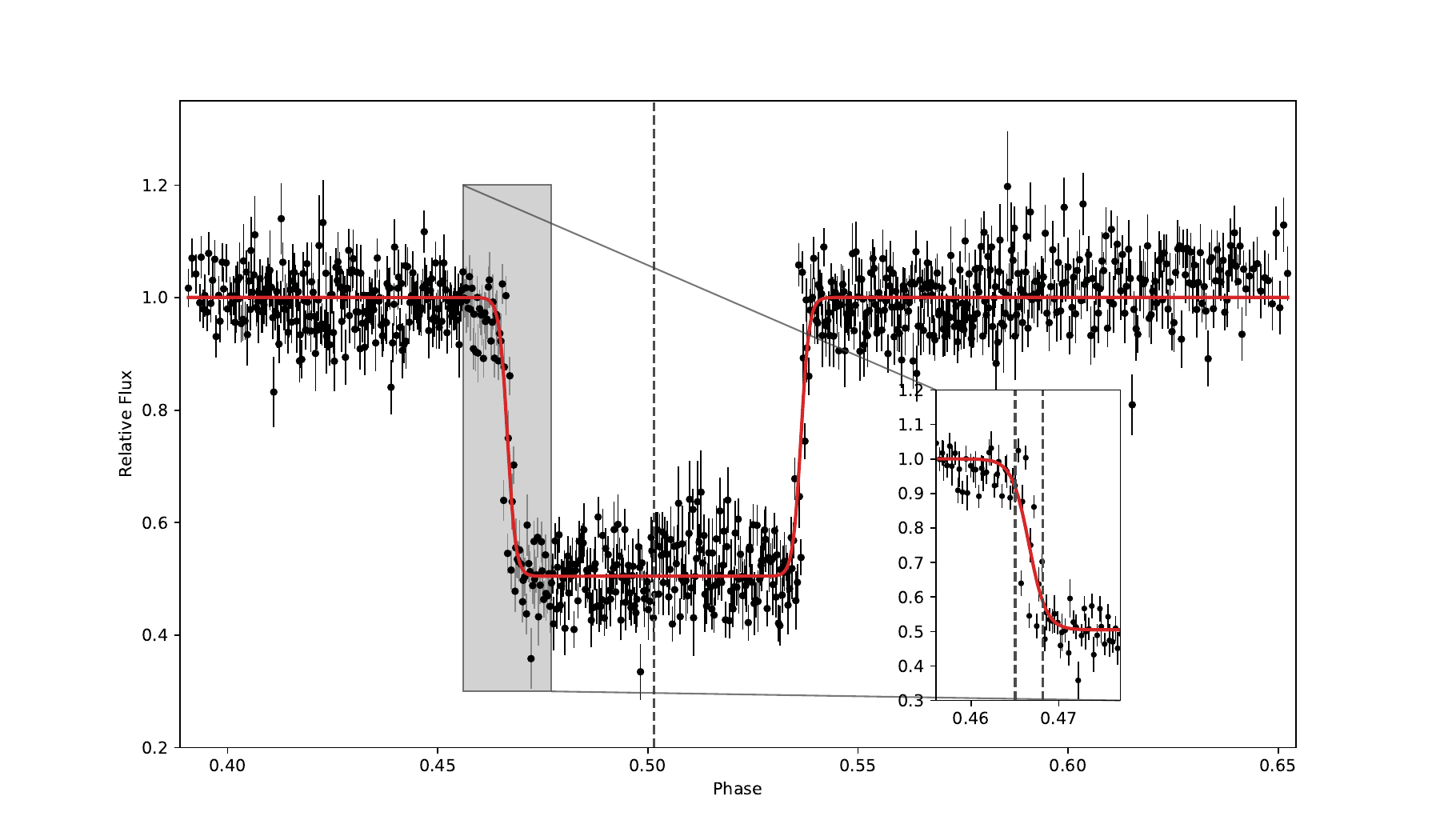}
	\caption{Fitting results of the light curve from the Liverpool Telescope. The black dots represent the data points, while the red curve shows the fitted model corresponding to the parameters listed in Table \ref{tab:coef}.}
	\label{fig:lt}
\end{figure}

The \texttt{curve\_fit} function from the \textit{scipy}\footnote{\url{https://scipy.org}} package was employed to fit the LT light curve using the least-squares method. The fitting results are illustrated in Figure \ref{fig:lt}, with the coefficients and their associated errors summarized in Table \ref{tab:coef}.

\begin{table}[t] 
    \centering
    \caption{The optimal fitting parameters for the Liverpool light curve data were obtained using Equation \ref{eq:eclip}.}
    \begin{tabular}{llll}
    \hline
        Name & Values & Description\\ \hline  
        k & 1027 $\pm$ 103 &  Rate of Change \\   
        c & 0.505 $\pm$ 0.004 & Phase at Mid-eclipse \\   
        d & 0.496 $\pm$ 0.005 & Depth of the Eclipse \\   
        $w_1$ & 0.0349 $\pm$ 0.0001 & Phase Shift \\ 
        s & 0.5015 $\pm$ 0.0001 & Flux offset \\ \hline
    \end{tabular}
    \label{tab:coef}
\end{table}

\begin{figure*}[t]
\gridline{\fig{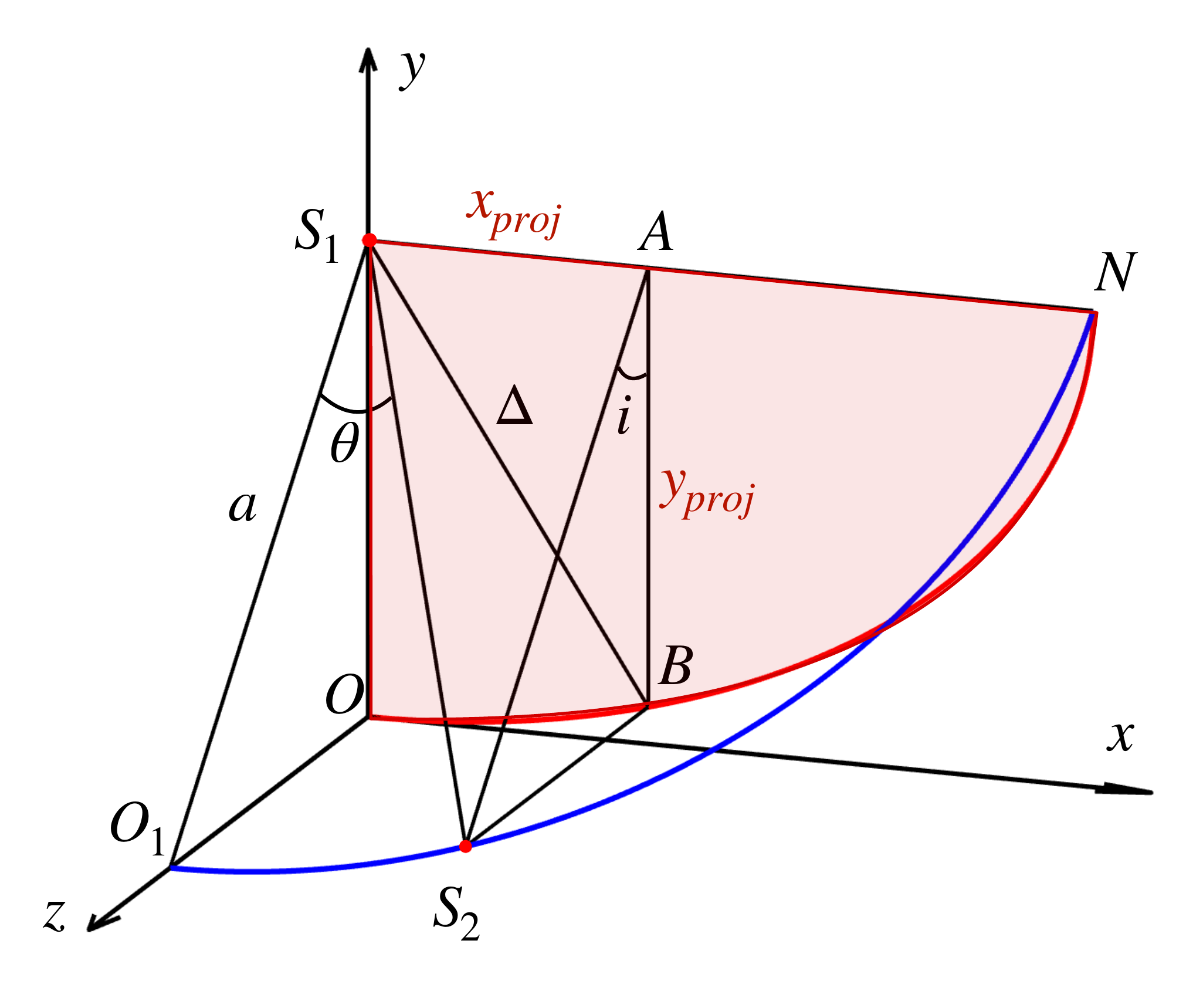}{0.4\textwidth}{(a)}
          \fig{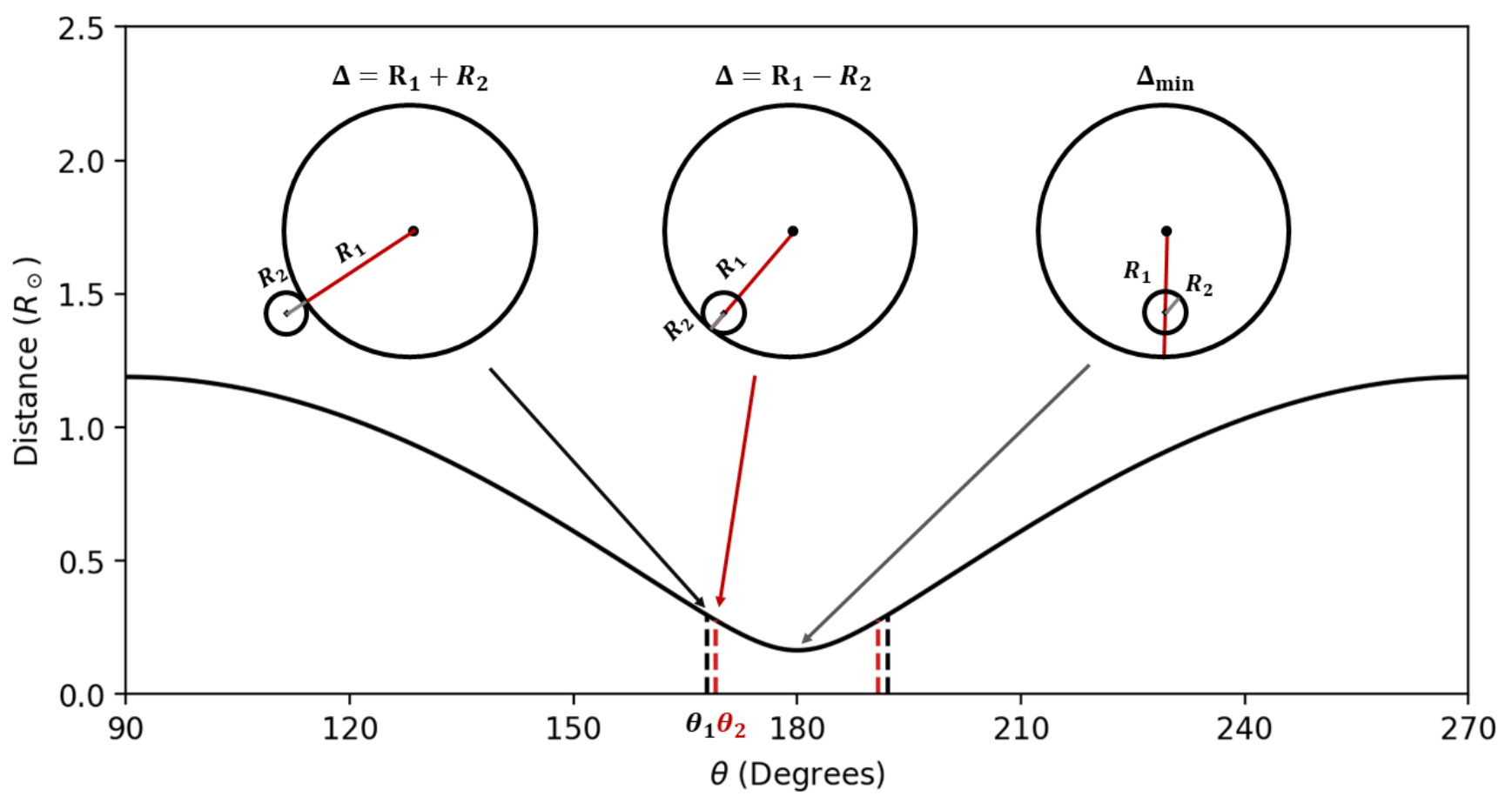}{0.6\textwidth}{(b)}
          }
\caption{\textbf{(a)} Schematic of geometric orbit. The blue curve represents the binary orbit, while the red-shaded region corresponds to the projection plane. The z-axis is aligned with the line of sight. S1 and S2 denote the main-sequence star and the white dwarf, respectively. 
		\textbf{(b)} The relationship between orbital phase \(\theta\) and the projected separation between the two components.
		Note that, for clarity in illustrating the relative positions of the two stars, the radius of the white dwarf has been moderately enlarged in this diagram.
\label{fig:theta_d}}
\end{figure*}


\subsubsection{Eclipse Modeling}\label{subsec:eclip_model}  


J1013+2724 was identified as a circular orbital system based on its orbital solution. The system exhibited eclipsing features, and the LT light curve showed a nearly constant flux outside the eclipse.
The geometric orbital model proposed by \citet{cherepashchuk2022close} was referenced to analyze the light curve of this binary system.
This geometric orbital model, combined with a binary model, facilitated the determination of key physical parameters, including the orbital inclination, stellar radii, and masses. Additionally, the combined model was integrated with MCMC simulations to explore the solution space that satisfies the eclipsing features of J1013+2724.

First, the parameter space was defined. It was divided into two parts. The first part included parameters obtained from previous data analyses, along with their uncertainties. These parameters consisted of the radial velocity amplitudes \(K_1\) and \(K_2\) of the two stars derived from velocity curve fitting, the orbital period \(P_{\text{orb}}\), and the photometric radii of the two components (\(R_1\) and \(R_2\)) obtained through SED fitting. Although the orbital inclination \(i\) was not precisely determined, the presence of eclipses, a short orbital period (\(\sim\)3 hours), and a significant radius difference between the two stars suggested a high inclination. Consequently, an initial estimate of \(70^\circ \sim 90^\circ\) was provided. 

The second part of the parameters was derived from known physical quantities (parameters from the first part) using a simplified binary star model. These parameters included the masses of the two stars (\(M_1\) and \(M_2\)) and the orbital semi-major axis (\(a\)). The mass function \(f(m)\) was calculated using the first set of parameters (\(K\), \(P_{\text{orb}}\)) 
\begin{equation}
	f(m)=\frac{m_2^2\sin^3i}{(m_1+m_2)^2}=\frac{P_{\text{orb}}\cdot K_1^3}{2\pi G} 
	\label{eq:fm} 
\end{equation} 
where \(G\) is the gravitational constant. Since the system follows a circular orbit, the mass ratio can be expressed as \(q=M_2/M_1=K_1/K_2\). Combined with Equation \ref{eq:fm}, the masses of the two stars were determined for each set of parameters. Once the masses were known, the orbital semi-major axis \(a\) was calculated using Kepler's third law 
\begin{equation}
	a=\left[\frac{G(m_1+m_2)P_{\text{orb}}^2}{4\pi^2}\right]^{1/3}
\end{equation}

To ensure that an eclipse occurs for each set of parameters, the orbital inclination was further screened. For each set of parameters, the minimum inclination angle \(i_{\text{min}}\) was calculated. This value corresponds to the scenario where the two stars are tangentially aligned in the line-of-sight projection plane, just avoiding an eclipse. The formula for \(i_{\text{min}}\) is \(i_{\text{min}}=\arccos[(r_1+r_2)/a]\). Any inclination angle \(i\) smaller than \(i_{\text{min}}\) was excluded from the parameter space. 

With these constraints, any parameter set within the space defines a complete geometric orbit. For each set, the projected distance between the two stars was calculated to determine the ingress, egress, and total duration of the eclipse. The geometric orbit of the binary system is illustrated in Figure~\ref{fig:theta_d}(a), where the blue line represents the actual orbit, and the red line shows the projected orbital plane. S1 and S2 denote the main-sequence star and the white dwarf, respectively. Their separation is \(a\), \(\theta\) is the orbital phase (with the midpoint of the eclipse set to zero phase), and \(i\) is the orbital inclination. The projected distance \(\Delta\) is given by 
\begin{equation}
	\Delta=\sqrt{x_{\text{proj}}^2+y_{\text{proj}}^2}
\end{equation}
where \(x_{\text{proj}}=a\sin(\theta)\) is the projected length in the x-direction, and \(y_{\text{proj}}=a\cos(\theta)\cos(i)\) is the projection in the y-direction. Figure~\ref{fig:theta_d}(b) shows how \(\Delta\) varies with orbital phase \(\theta\). 
The ingress begins at \(\Delta=R_1+R_2\) and ends at \(\Delta=R_1-R_2\), with the midpoint of the eclipse corresponding to \(\Delta_{\text{min}}\).
Using these relationships, the eclipse characteristics were readily calculated. 

\begin{figure*}[ht!]
	\includegraphics[width=0.9\textwidth]{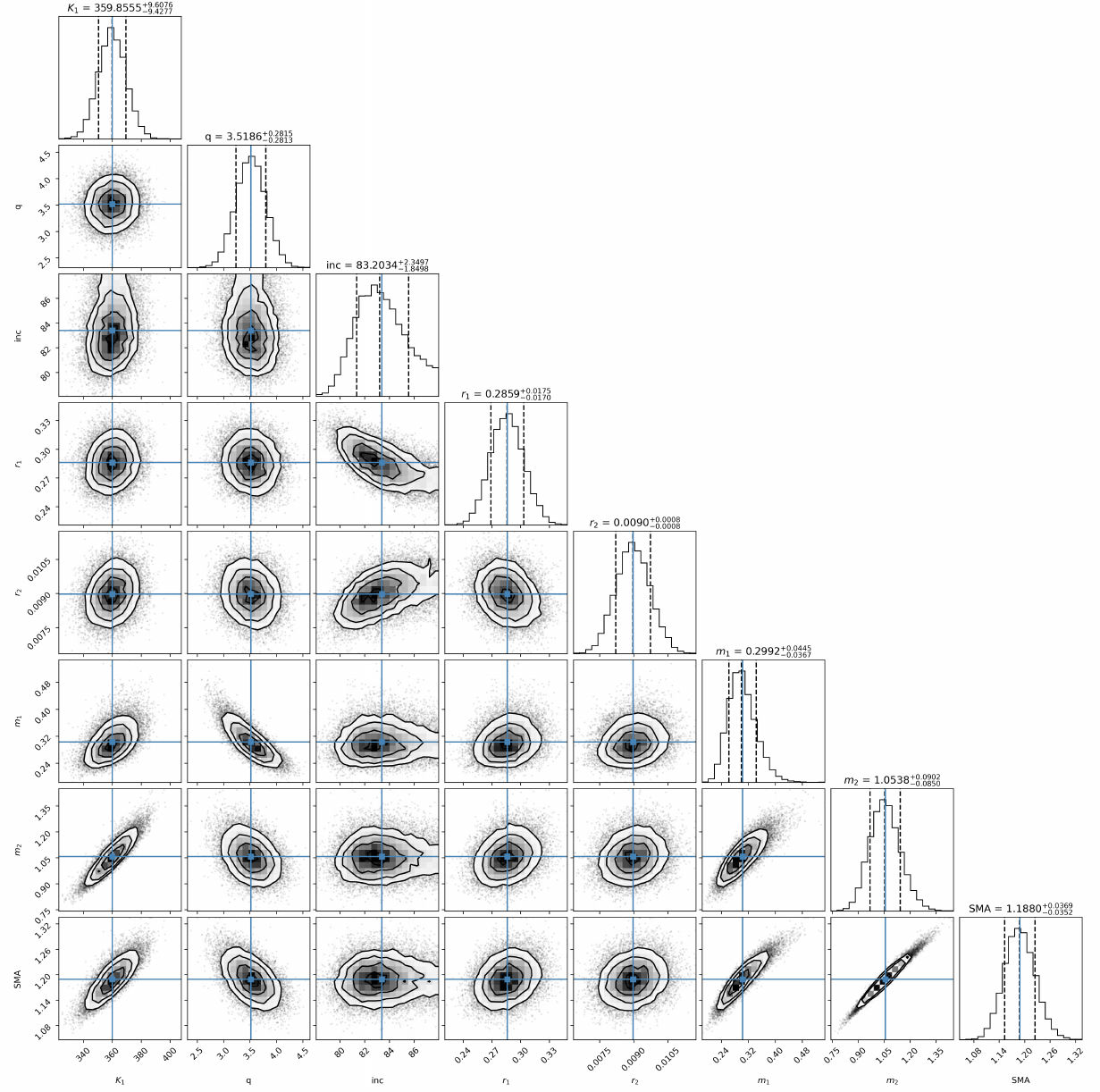}
	\caption{Corner plot derived from the eclipsing model.}
	\label{fig:eclip_corner}
\end{figure*}

The eclipse timing information derived from the geometric orbital model for various parameter combinations was compared with the eclipse durations obtained from fitting the LT light curve data in Section 3.4.1 (\(t_{\text{ingress}}=t_{\text{egress}}=36\pm3 \, \text{s}\), \(\theta_{\text{eclipse}}=0.073\)). Using this comparison, the solution space consistent with observational data was identified. 
We applied an MCMC algorithm to randomly sample the binary system's parameter space, generating 100,000 samples. For each sample, the ingress timing and eclipse duration were calculated. The \(\chi^2\) value between the calculated eclipse characteristics and the fitted observational results was then evaluated. The acceptance or rejection of each parameter sample was determined by comparing the corresponding probability of the \(\chi^2\) distribution with a randomly generated number between 0 and 1. Samples with probabilities lower than the random number were rejected, while those with higher probabilities were accepted. 
The final MCMC-derived solution space for the binary system is illustrated in Figure~\ref{fig:eclip_corner}. The results for each physical parameter, along with their uncertainties, are summarized in Table~\ref{tab:results}. 

To better understand the sources of parameter uncertainties, we analyzed how radial velocity measurement errors propagate through the mass function. To account for this, we calculated the mass errors caused by \(K_1\) and \(K_2\) (or the mass ratio $q$) in addition to statistical uncertainties. The resulting uncertainties are 0.09 \(M_\odot\) for the main-sequence star and 0.18 \(M_\odot\) for the white dwarf.

\subsection{Theoretical Light Curve} \label{subsec:ztf}  

By the end of the Section \ref{subsec:eclip}, the key stellar parameters, including masses and radii, had been reliably determined. To validate the stellar parameters derived in Section \ref{subsec:eclip}, theoretical light curves for the ZTF g and r-bands, as well as the TESS band, were generated using the Wilson-Devinney code \citep{wilson1971realization,2020ascl.soft04004W}, a widely used tool for modeling monochromatic light curves of close eclipsing binary systems. 
In the phase-folded ZTF g and r-band, as well as TESS light curves (Figure \ref{fig:rv_fit}), the regions outside the eclipse are primarily influenced by ellipsoidal effect. However, this effect is not evident in the LT light curve (Figure \ref{fig:lt}), as the LT data only cover the orbital phases corresponding to the eclipse and its immediate surroundings.
 These light curves were not directly fitted using the Wilson-Devinney code to avoid overfitting. Since the eclipses provide sufficient constraints on key parameters, incorporating the entire light curve for fitting could unnecessarily emphasize regions outside the eclipse. This would reduce the relative weight of the eclipse data, resulting in suboptimal fits to the critical eclipse features. Thus, parameters derived from the eclipse were used as a foundation for generating theoretical light curves, allowing for a robust and independent comparison with the observed data.

The extinction values for the respective bands were adopted as \(A_g = 0.0350 \pm 0.0118 \, \text{mag}\), \(A_r = 0.0250 \pm 0.0084 \, \text{mag}\), and \(A_{\text{TESS}} = 0.0202 \pm 0.0068 \, \text{mag}\). 
Additional data cleaning was applied to the TESS observations. Due to the relatively long exposure time of TESS (10 minutes) compared to the orbital period, the eclipse minima were almost indistinguishable. To minimize contamination, data points within the phase range of \(0.5 \pm 0.04\) were excluded. The phase-folded TESS light curve was then fitted with a sinusoidal model, and outliers with residuals greater than \(3\sigma\) were removed using a sigma-clipping method. This process was repeated iteratively to ensure robust cleaning of the data.

Limb darkening, gravity darkening, and reflection effects were considered in our analysis. The limb-darkening coefficients were internally determined by the Wilson-Devinney code based on the square-root law. The gravity darkening (GR) and reflection coefficients (ALB) were set based on the stellar envelope conditions. For the main-sequence star, whose outer layers are likely convective, GR was set to 0.3 and ALB to 0.5. For the hotter white dwarf, both GR and ALB were set to 1. Additionally, as discussed in Section \ref{subsec:emission_line}, M stars may exhibit stellar activity of varying intensity, which could lead to differences in observational results across bands. To account for this, we adjusted the effective temperatures of both stars within the error margins to better match the light curves in different bands.

The theoretical light curves are shown in the lower panels of Figure \ref{fig:rv_fit}. The light curves generated by the WD code closely match the ZTF observations in the g- and r-bands. The high level of agreement between the simulated and observed data confirms the reliability of the parameters derived from the eclipsing binary model.


\section{Results and Discussion} \label{sec:result}  

\subsection{Mass, radius and temperature} \label{subsec:para_discussion}

\begin{figure*}[ht!]
	\centering
	\includegraphics[width=0.8\textwidth]{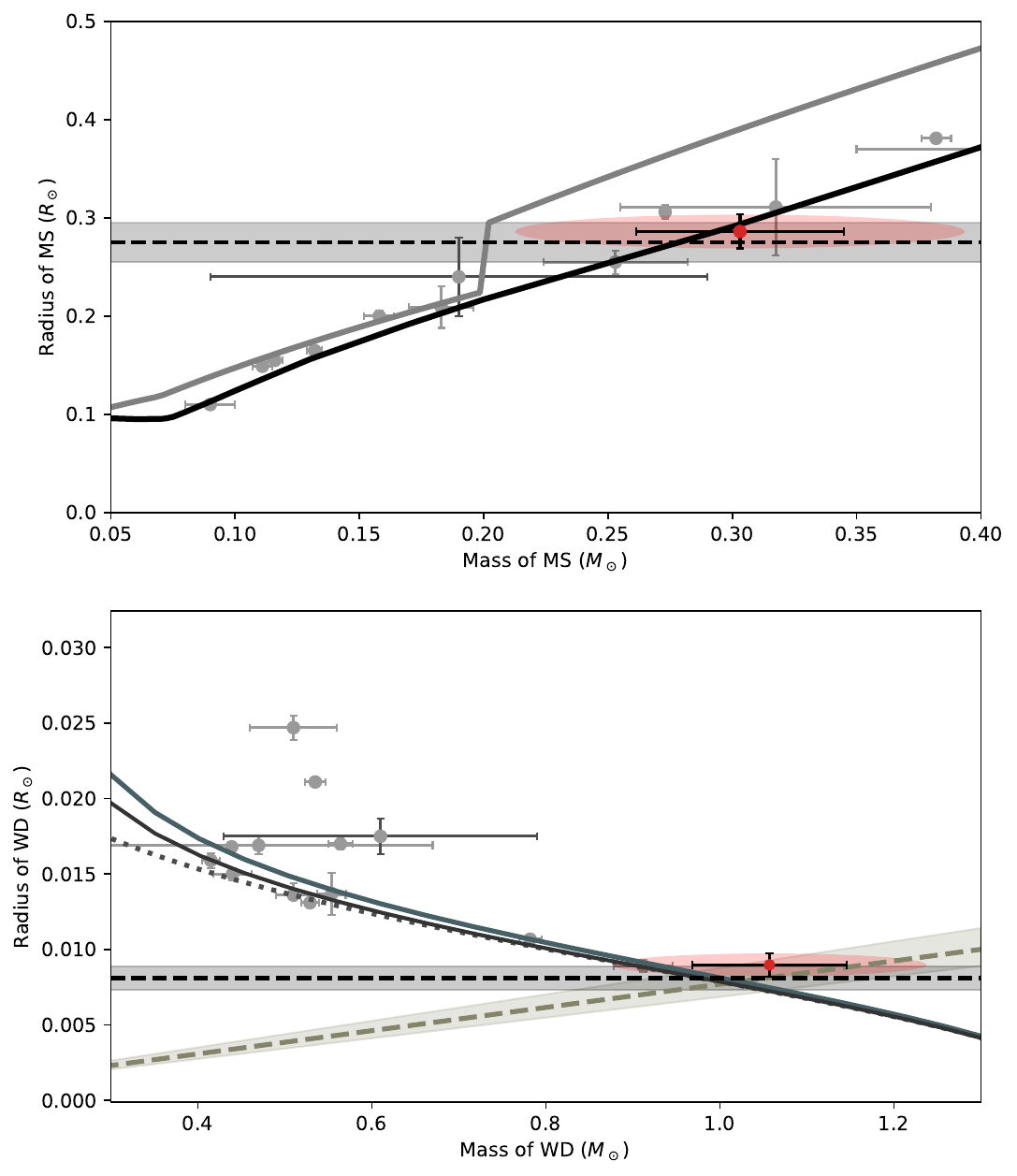}
	\caption{The mass-radius relations for the main-sequence star (top) and white dwarf (bottom) are shown. In the upper panel, the black solid line represents the 1 Gyr isochrone from \citet{baraffe2015new}, while the gray solid line shows the mass-radius relation for M stars in CVs as reported by \citet{knigge2011evolution}. In the lower panel, the solid lines correspond to the theoretical mass-radius relations of a 15,600 K white dwarf interpolated from the models of \citet{fontaine2001potential}, with a thick hydrogen envelope (\(q_H=10^{-4}\), cyan) and a thin hydrogen envelope (\(q_H=10^{-10}\), black). The dotted line represents the zero-temperature mass-radius relation of Eggleton, as cited in \citet{verbunt1988mass}. The brown dashed line corresponds to Equation \ref{eq:vg}, where the gravitational redshift is \(V_g = 82.9 \pm 10.2 \, \text{km/s}\). The black dashed line and the gray shaded region indicate the photometric radii derived from SED fitting and its associated uncertainties.  The red dot in the figure represents J1013+2724, with the black error bars indicating statistical uncertainties. The ellipse's major axis reflects the mass errors propagated from the uncertainties in radial velocity amplitudes. The gray points denote other direct mass-radius measurements from eclipsing PCEB systems \citep{parsons2010orbital,parsons2012shortest,parsons2012precision,parsons2012accurate,parsons2016crowded, pyrzas2012post,van2017ptf1}.} 
	\label{fig:mr_relation}
\end{figure*}

In this section, we compare the masses, radii, and effective temperatures of the two components obtained through different methods, and assess their consistency with previous studies of this system. The results of this work are summarized in Table \ref{tab:results}, with the masses and radii of the components plotted in Figure \ref{fig:mr_relation}.

The mass and radius of the white dwarf, determined through eclipse analysis, are \(1.05 \pm 0.09 \, M_\odot\) (with an uncertainty of  \(0.18 \, M_\odot\) propagated from the mass ratio) and \(0.00897 \pm 0.00077 \, R_\odot\), respectively. These values are consistent within uncertainties with the mass-radius relation for a \(15661 \, \text{K}\) CO white dwarf presented by \citet{fontaine2001potential} (Figure \ref{fig:mr_relation}), but are slightly larger than the photometric radius of \(0.0081 \pm 0.0003 \, R_\odot\) obtained from SED fitting. Using the effective temperature and surface gravity, \citet{rebassa2012post} determined a white dwarf mass of \(1.1 \pm 0.023 \, M_\odot\) and a radius of \(0.00698 \pm 0.00027 \, R_\odot\) based on the tables of \citet{bergeron1995photometric}. Our derived mass is slightly lower, and our radius is slightly larger than their results. While there are differences, the results are broadly consistent and may arise from the methods used to derive the parameters.
Although the mass of the white dwarf in this system lies within the range of both CO-core and ONe-core white dwarfs (within measurement uncertainties), no spectral lines of heavy elements (e.g., Mg II) were detected. Additionally, for the same mass, ONe-core white dwarfs have relatively smaller radii, whereas the observed radius of this white dwarf is slightly larger than the theoretical estimate for CO-core white dwarfs. Hence, it is more likely that the white dwarf in this system is a CO-core white dwarf.

The main-sequence star’s mass and radius, derived from eclipse analysis, are \(0.299 \pm 0.045 \, M_\odot\) (with an uncertainty of  \(0.09 \, M_\odot\) propagated from the mass ratio) and \(0.286 \pm 0.018 \, R_\odot\), respectively. Its radius slightly exceeds the photometric radius of \(0.275 \pm 0.022 \, R_\odot\), obtained through SED fitting, as shown in Figure \ref{fig:mr_relation}. Based on the Sp-M-R relation from \citet{rebassa2007post}, \citet{rebassa2012post} reported a mass of \(0.319 \pm 0.09 \, M_\odot\) and a radius of \(0.326 \pm 0.096 \, R_\odot\) for this M4-type star. Both values are larger than the results obtained in this work. The discrepancies may arise from their use of empirical relations, which are related to the precision of M star parameter measurements. In contrast, our results were obtained through light curve analysis, which provides more detailed information and better constraints from the eclipse.

The effective temperatures derived from SED fitting are \(T_{\rm eff, 1} = 3196 \pm 83 \, \mathrm{K}\) for the M star and \(T_{\rm eff, 2} = 15606 \pm 394 \, \mathrm{K}\) for the white dwarf. These values are slightly lower than, but consistent within uncertainties with, the spectroscopically determined temperatures of \(T_{\rm eff, 1} = 3307.5 \pm 39.5 \, \mathrm{K}\) and \(T_{\rm eff, 2} = 15661 \pm 973 \, \mathrm{K}\).
The white dwarf's effective temperature from our spectral analysis aligns with the value reported by \citet{rebassa2012post} (\(T_{\rm eff} = 16526 \pm 277 \, \mathrm{K}\)) and the result derived by \citet{parsons201514} (\(T_{\rm eff} = 15601 \, \mathrm{K}\)).
Minor numerical discrepancies in the white dwarf's temperature likely result from differences in analysis methods and data sources: \citet{parsons201514} derived the temperature from the system's u-g color, while \citet{rebassa2012post} obtained it by fitting SDSS spectra.

\subsection{Production of the emission-line on the M star} \label{subsec:emission_line}

The spectrum of J1013+2724 exhibits several prominent emission lines. As shown in Figure \ref{fig:cospec}, the white dwarf's \(H\alpha\) absorption line is barely discernible and is replaced by emission lines that shift with orbital phase. Similarly, in the blue end of the spectrum, narrow emission cores are observed in the centers of the white dwarf's H Balmer lines, such as  \(H\beta\), which also follow the motion of the main-sequence star.

These emission lines could originate from two possible mechanisms: (1) chromospheric activity on the main-sequence star, as M dwarfs are typically active low-mass stars with strong magnetic fields and chromospheric emissions, or (2) the irradiation effect, where the white dwarf heats the atmosphere of the main-sequence star, causing additional radiation from the irradiated surface. In the second case, the velocities of the emission lines do not track the stellar center of mass but rather the center of light of the irradiated hemisphere. These lines are expected to be strongest when the heated face directly faces the observer. This can be tested by measuring the radial velocity semi-amplitudes and the equivalent widths (EWs) of the lines at different orbital phases.

By performing Gaussian fits to the normalized single-exposure spectra, we measured the radial velocities and EWs of the \(H\beta\) core emission, \(H\alpha\) emission, and Na I 8183/8194 absorption doublet (used as a reference). The RV semi-amplitudes were determined to be \(K_{H\beta}=358 \pm 8 \, \mathrm{km/s}\), \(K_{H\alpha}=335 \pm 11 \, \mathrm{km/s}\), and \(K_{NaI}=346 \pm 11 \, \mathrm{km/s}\), which are consistent within uncertainties. The slight differences might result from variations in optical depth among the lines \citep{parsons2010precise}. Assuming the emission lines are due to the irradiation effect, the RV amplitude of the main-sequence star's center of mass should be significantly larger than the measured values of the \(H\alpha\) line. After applying the  \(K_\mathrm{sec}\)  correction from \citet{parsons2010precise}, the corrected  \(K_{H\alpha}\)  amplitude reaches \(412 \, \mathrm{km/s}\) (with \(f=0.42\)), significantly exceeding the RV amplitude of the Na I doublet. Thus, the emission lines are unlikely to originate from irradiation. Furthermore, an analysis of the equivalent widths as a function of orbital phase revealed no significant enhancement at zero phase (the moment when the heated face directly faces the observer). A correlation test confirmed that EWs exhibit no significant dependence on orbital phase. These results suggest that irradiation in this system is relatively weak, and the emission lines are more likely caused by stellar activity on the M dwarf rather than the irradiated surface.
Stellar magnetic activity indicates that the magnetic field still exists. However, when M-type stars become fully convective, the increase in surface magnetic complexity \citep[as suggested by][]{Taam1989} may cause the system to continue losing angular momentum at a slower rate via magnetic braking, rather than the dynamo itself shutting down.



\subsection{Gravitational Redshift of WD}\label{subsec:vg_mr}  

According to general relativity, light emitted from a white dwarf undergoes gravitational redshift, losing energy and shifting to longer wavelengths as it escapes the star's gravitational field. This redshift is given by:
\begin{equation}\label{eq:vg}
v_g=\frac{\delta \lambda\cdot c}{\lambda_0}=GM/cR    
\end{equation}
where \(\delta \lambda\) is the wavelength shift, \(\lambda_0\) is the rest wavelength, G is the gravitational constant, M is the white dwarf's mass, c is the speed of light, and R is its radius. This equation connects the observed gravitational redshift to the white dwarf's mass and radius.

To begin, in Section \ref{subsec:rv_fit}, we separated the effects of Doppler shifts and gravitational redshift by analyzing the radial velocity of the main-sequence star. This analysis yielded a gravitational redshift of \(v_g = 82.9 \pm 10.2 \)km/s.
Next, in Section \ref{subsec:eclip}, we determined the white dwarf's mass and radius through eclipse modeling. The results, \(M = 1.05 \pm 0.09 \, M_\odot\) and \(R = 0.00897 \pm 0.0077 \, R_\odot\), correspond to a gravitational redshift of \(v_g = 75.0 \pm 9.1\) km/s when applied to Equation \ref{eq:vg}. These findings provide an initial comparison point between observed and modeled parameters.
Furthermore, by applying the white dwarf cooling model from \citet{fontaine2001potential}, we interpolated the theoretical mass-radius relationship specific to white dwarfs. This approach predicted a radius of \(R = 0.0076 \pm 0.0011 \, R_\odot\) for a white dwarf with \(M = 1.05 \pm 0.09 \, M_\odot\). Substituting this radius into Equation \ref{eq:vg} provided a theoretical gravitational redshift of \(v_g = 88.7 \pm 14.7\) km/s.

In conclusion, the observed and theoretical gravitational redshift values agree within their respective uncertainties. This consistency not only supports the alignment between our models and observations but also demonstrates the robustness and reliability of our analytical methods.

\subsection{Is J1013+2724 crossing the period gap as a dCV?} \label{subsec:history}

Based on the analysis above, J1013+2724 is a WD+MS binary system located within the period gap of cataclysmic variables, with an orbital period of 185.82 minutes. This section explores two possible evolutionary scenarios for this system: (1) whether it directly evolved to its current state as a detached post-common-envelope binary, never experiencing mass transfer (thus not entering the CV phase), or (2) whether it has already evolved into a CV and is currently crossing the period gap as a detached CV.

If J1013+2724 directly formed as a PCEB from the CE phase, the white dwarf’s temperature would not be affected by material accretion. According to the cooling models of CO-core white dwarfs \citep{fontaine2001potential}, an effective temperature of approximately 15661 K corresponds to a cooling age of about 0.58 Gyr. If the system is a dCV, the white dwarf’s cooling age would be extended due to heating from accreted material. Given that the orbital period would decrease from 191 minutes to 185 minutes in approximately 0.05 Gyr through angular momentum loss via gravitational radiation \citep{faulkner1971ultrashort}. 
Furthermore, after detaching from the Roche lobe, the relaxation timescale of the main-sequence star is $\tau_{relax} \sim 0.03 \, \text{Gyr}$ \citep{king1995consequential}, which is shorter than the time the system has already spent in the period gap. Therefore, the main-sequence star should have returned to thermal equilibrium, with no significant radius inflation, consistent with the theoretical mass-radius relation for 1 Gyr main-sequence stars shown in Figure \ref{fig:mr_relation} \citep{baraffe2015new}.
However, CV evolutionary models predict that white dwarfs at the upper boundary of the period gap should have effective temperatures in the range of 23000-30000 K \citep{van2017ptf1}. Cooling from these temperatures to the observed 15661 K would require approximately 0.39-0.50 Gyr, which is significantly longer than the time the system would have spent in the period gap. This discrepancy suggests that if J1013+2724 is a dCV, the current white dwarf temperature should be higher than observed.

The system contains a massive white dwarf (\(\sim 1.05 M_\odot\)), consistent with the mass characteristics of white dwarfs in dCVs but rare among PCEBs, as suggested by the results of binary population models presented in \citet{zorotovic2016detached}.
 \citet{parsons2016crowded} studied a similar system, QS Vir, which hosts a massive white dwarf (0.782 $M_\odot$) with an orbital period of 3.6 hours. In QS Vir, the donor star also lacks the 30\% radius inflation predicted by \citet{knigge2011evolution}. Thus, they proposed that QS Vir is more likely a pre-CV than a detached CV (either hibernating or genuinely detached).

In conclusion, this system is more likely a PCEB, or pre-CV, as the white dwarf appears cooler than expected. However, the possibility of it being a dCV cannot be entirely ruled out. Currently, observational samples of detached systems located within or near the period gap are relatively limited. In particular, the scarcity of eclipsing binary samples makes it challenging to accurately characterize the statistical distribution of key parameters, such as stellar mass and radius. A larger sample size could better reveal the shared characteristics and differences among various types of binary systems. Specifically, confirming the presence of massive white dwarfs in PCEBs would enhance the distinction between PCEBs and dCVs, thereby advancing and refining CV evolution theories.

Regardless of the evolutionary history described in this section, the subsequent evolution of the system remains the same. The system loses angular momentum due to gravitational radiation, causing orbital contraction. It is predicted that after 0.27 Gyr, the orbital period will decrease to approximately 155 minutes, marking the lower boundary of the period gap. At this point, the donor is expected to refill its Roche lobe and remain filled, ending the separation and initiating mass transfer. The system then continues its evolution as a cataclysmic variable.


\section{ACKNOWLEDGEMENTS} \label{sec:ack}  
This work is supported by the National Natural Science Foundation of China (12273056, 12090041, 11933004) and the National Key R\&D Program of China (2019YFA0405002, 2022YFA1603002). Y.HL and B.ZR acknowledge support from the National Key R\&D Program of China (Grant No. 2023YFA1607901).

This work has made use of data from the Guoshoujing Telescope (the Large Sky Area Multi-Object Fiber Spectroscopic Telescope, LAMOST). LAMOST is operated and managed by the National Astronomical Observatories, Chinese Academy of Sciences (\href{http://www.lamost.org/public/?locale=en}{http://www.lamost.org/public/?locale=en}). Funding for the LAMOST has been provided by the National Development and Reform Commission.
This work uses data from the ESA Gaia mission, processed by the Gaia Data Processing and Analysis Consortium (DPAC) and funded by MLA member institutions. Gaia mission information is available at \href{https://www.cosmos.esa.int/gaia}{https://www.cosmos.esa.int/gaia}, with data access at \href{https://archives.esac.esa.int/gaia}{https://archives.esac.esa.int/gaia}. 

We acknowledge use of the VizieR catalog access tool, operated at CDS, Strasbourg, France, and of Astropy, a communitydeveloped core Python package for Astronomy (Astropy Collaboration, 2013).
We also acknowledge data from the Zwicky Transient Facility (ZTF, \href{https://www.ztf.caltech.edu}{https://www.ztf.caltech.edu}), funded equally by the U.S. National Science Foundation and an international consortium of universities and institutions. Additionally, we thank the Liverpool Telescope (\href{https://telescope.livjm.ac.uk}{https://telescope.livjm.ac.uk}), operated by the Astrophysics Research Institute of Liverpool John Moores University, for its valuable observational data.


\bibliographystyle{aasjournal}
\bibliography{J1013_2724}

\end{document}